\documentclass[twocolumn,letter]{jpsj2} 

\title{
Possibility of Solid-Fluid Transition in Moving Periodic Systems
}

\author{
Tomoaki \textsc{Nogawa}\thanks{E-mail address: 
nogawa@presto.phys.sci.osaka-u.ac.jp, 
Present address:
Division of Physics, Graduate School of Science, Hokkaido University, 
Kita 10-jo Nisi 8-tyome, Sapporo 060-0810.},
Hajime \textsc{Yoshino}\thanks{E-mail address:
yoshino@ess.sci.osaka-u.ac.jp,
Present address:
Laboratoire de Physique Th{\'e}orique  et Hautes {\'E}nergies, Jussieu,
5\`eme {\'e}tage,  Tour 25, 4 Place Jussieu, 75252 Paris Cedex 05, France.},
and Hiroshi \textsc{Matsukawa}$^{1}$\thanks{E-mail address: 
hm@phys.aoyama.ac.jp}
}

\inst{
School of Science, Osaka University,
Machikaneyama 1-1, Toyonaka, Osaka 560-0043 \\
$^{1}$
College of Science and Engineering, Aoyama Gakuin University, \\
Fuchinobe 5-10-1, Sagamihara, Kanagawa 229-8558 \\
}

\abst{
The steady sliding state of periodic structures 
such as charge density waves and flux line lattices is numerically studied 
based on the three dimensional driven random-field XY model.
We focus on the dynamical phase transition 
between plastic flow and moving solid phases controlled 
by the magnitude of the driving force.
By analyzing the connectivity of comoving clusters, 
we find that they percolate the system 
under driving forces larger than a certain critical force 
within a finite observation time.
The critical force, however, logarithmically diverges 
with the observation time, 
i.e., the moving solid phase exists only within a certain finite time, 
which exponentially grows with driving force.
}

\kword{CDW, flux line lattice, plastic flow, Bragg glass, random field, XY model}

\begin{document}

\maketitle


The collective transport phenomena of condensed matter
with random pinning attract much attention
from the viewpoints of solid state physics, nonlinear dynamics
and statistical mechanics.
There are numerous systems which belong to this class of dynamics,
e.g., charge density waves (CDWs) \cite{Gruner88},
flux line lattices (FLLs) \cite{Blatter94},
colloidal lattices \cite{Reichhardt02} 
and  Wigner crystals \cite{Williams91,Shirahama04}.
These systems  have spatial periodicity in the absence of random 
pinning, which modifies the periodic order and pins the system.
Under a driving force larger than a certain threshold value, 
the system starts moving and shows highly nonlinear conduction.
This is the depinning transition, which has been  investigated extensively.
Recently another attractive topic of these systems has been 
the dynamical phase transition between two nonequilibrium steady states 
above the depinning threshold field \cite{Koshelev94,Balents95}.
In the ``ordered phase" the local DC velocity is uniform and the 
spatial periodicity has a quasi long range order.
Such a state in FLL systems is called moving Bragg glass 
\cite{Balents98, Doussal98}.
In the ``disordered phase", which is often called the ``plastic flow phase", 
the motion is spatially nonuniform and the periodic order is destroyed.
This transition is considered to be induced by a change 
in driving force or pinning strength.
Although there are a lot of experimental studies 
\cite{Higgins96,Pardo98} and numerical simulations 
\cite{Olson98,Dominguez99,Kolton99,Kolton02,Chen03}
that show an evidence of a dynamical melting transition 
between moving Bragg glass and fluid phases,
the existence of such an ordered phase is still an unanswered question.

In such discussions, it is assumed implicitly that two types of 
order, namely, the spatial periodic order and the local DC 
velocity order, are established simultaneously. 
To date, the former has been mainly discussed.
These two orders, however, are independent in principle,
e.g., the ``moving glass state", in which the spatial periodicity is 
destroyed but frozen in time so the local DC velocity is uniform, is 
possible.
In this article we focus on the dynamical property, i.e., the 
uniformity of the local DC velocity.
It is closely related to the plastic deformation because at the 
boundary between the domains that have different velocities,
the local strain increases with time and tearing occurs.
We discuss the phase transition 
between the ``plastic flow phase" and the ``moving solid phase", 
which are distinguished by the existence of the local DC velocity order.


We perform numerical simulations based 
on the three dimensional driven random-field XY model 
\cite{Strogatz88,Huse97,Kawaguchi99,Nogawa03}.
It is a modified version of the intensively investigated elastic 
manifold model,
such as the Fukuyama-Lee-Rice model for CDWs
\cite{Fukuyama78,Lee79}.
The density of a periodic structure is expressed as 
$
\rho(\mathbf{r},t) = \sum_{\mathbf{q}}
\rho_{\mathbf{q}} \cos
\big(
\mathbf{q} \cdot [
\mathbf{r}-\mathbf{u}(\mathbf{r},t)
]
\big)
$.
Here, $\mathbf{q}$'s are fundamental reciprocal lattice vectors.
Higher harmonics are ignored here.
$\mathbf{u}(\mathbf{r},t)$ is a deformation field
and the phase field 
$\theta_\mathbf{q}(\mathbf{r},t)=\mathbf{q\cdot u}(\mathbf{r},t)$ 
is often employed as a degree of freedom.
In this article, only the case of a single phase field, 
which denotes the driving direction component of deformation, is treated.
It is sufficient in the case 
that periodicity is one dimensional as CDWs in NbSe$_3$.
We consider, however, that the essential feature of the dynamics of
higher dimensional periodic systems such as FLLs
would be captured.
The periodicity of the structure is related to the phase coherence 
and experimentally observable transport quantities,
such as electric current for CDWs and voltage drop for FLLs,
are proportional to the phase velocity
$\dot{\theta}(\mathbf{r},t)$.

The elastic manifold model treats internal interaction 
using the elastic energy $\int d\mathbf{r} (\nabla \theta)^2$, 
which becomes a harmonic coupling,
$\sum_{\langle i,j \rangle} (\theta_i - \theta_j)^2/2$,
in a lattice model.
In order to treat plastic deformation,
we replace this harmonic coupling
with a sinusoidal one,
$1- \cos( \theta_i - \theta_j )$.
They are equivalent in the limit where phase 
differences become zero.
Here, the indices $i$'s denote semi-macroscopic domains fixed in the space
in which phase coherence is always held.
This sinusoidal coupling induces maximum restoring force, i.e., yield stress 
and allows plastic deformation, so-called phase slip.
Phase slip is a process in which the phase difference 
between neighboring domains increases or decreases by $2\pi$.
It results in no change in coupling energy.

The overdamped equations of motion for the phases of domains, 
$\theta_i$'s, are as follows.
\begin{equation}
\dot{\theta}_i = -\frac{J}{z} \sum_{j} \sin(\theta_i - \theta_j)
- \sin(\theta_i - \beta_i) + f
\label{eq:eom}
\end{equation}
We choose the units that both of a pinning strength
and a dissipation coefficient equal unity.
The first term on the right hand side refers to the interaction
with neighboring $z$ domains.
The second term denotes the a random pinning force 
and $\beta_i$'s are given as uniform random numbers between 0 and $2\pi$.
$f$ is a uniform driving force.

Strogatz {\it et al.} analyzed this model by the mean field approximation
and found a discontinuous transition
by changing the external field \cite{Strogatz88}.
There are three regimes,
a pinned static state and a homogeneously moving state, 
and a bistable regime between the two.
Huse performed numerical simulations of this model
in three dimensions \cite{Huse97}.
He investigated the Lyapunov exponent and velocity coherence
and found a transition between spatially uniform temporally regular motion 
and nonuniform chaotic motion by changing pinning strength.
These motions are related to
moving solid and plastic flow, respectively.
In this article, we analyze the $f$ dependence of the dynamics 
of this model systematically 
paying special attention to the dependences 
on system size and observation time.

We numerically solve eq. (\ref{eq:eom})
by the fourth-order Runge-Kutta method.
The discretized time step is set at $2\pi/8(J+f)$.
The domains are put regularly on 
the simple cubic lattice in three dimensions.
We call this unit ``site" instead of ``domain" hereafter.
A periodic boundary condition is imposed.
There are two independent parameters, coupling constant $J$
and driving force $f$.
In this article we show mainly the results for $J=1.0$. 
All phases are set at the unique value in the initial state.
Physical quantities are calculated
after some precursor running (typical time is 12900)
for the relaxation to the steady state.
An initial state is sometimes substituted
by the final state for the simulation with a slightly larger $f$.
Thermal fluctuation is not taken into account.
Simulations are performed with some samples 
that have different set of $\beta_i$'s. 
The numbers of samples whose linear sizes are 16,32,64 and 128
are 32,16,8 and 4, respectively.

\begin{figure}
\begin{center}
\includegraphics[trim=0 0 0 0,scale=0.38,clip]{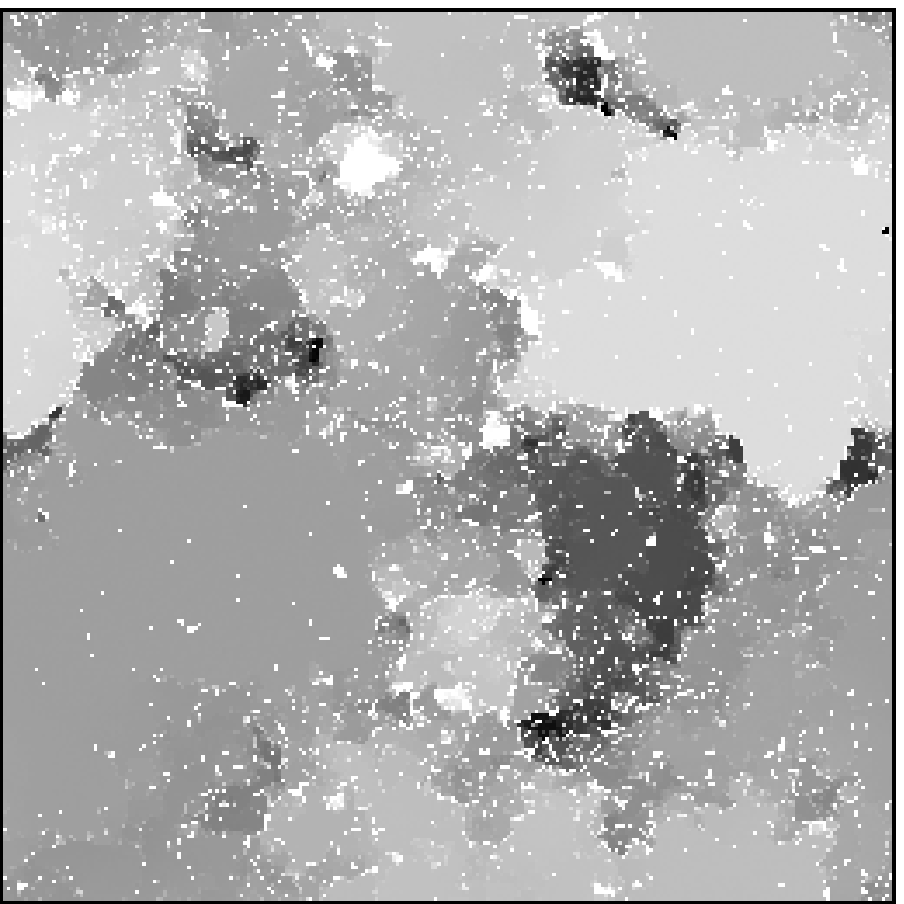}
\hspace{0.2cm}
\includegraphics[trim=0 0 0 0,scale=0.38,clip]{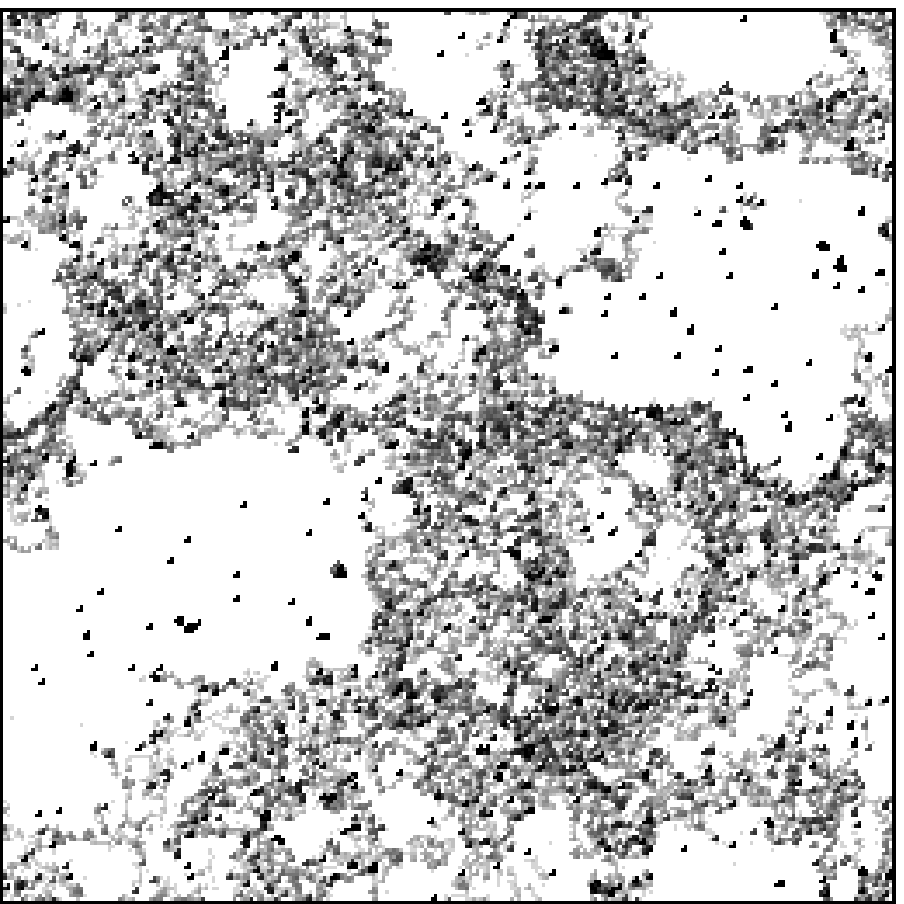}
\end{center}
\vspace{-2mm}
\caption{\label{fig:profile}
Spatial configuration of $\omega_\mathrm{DC}$ on site (left)
and $\Delta \omega_\mathrm{DC}$ on bond (right)
in a two dimensional sample.
$\omega_\mathrm{DC}T/2\pi$ is plotted
with gray scale from 2730 to 2820 as color changes from white to black 
and $\log(|\Delta \omega_\mathrm{DC}| T/2\pi)$ changes
from $\log0.25$ to $\log1000$.
Connected bonds are plotted in white.
Result of two dimensional system with $256^2$ sites,
$T \approx 25700$, $J$=1.0 and $f$ = 1.0 ($f_c(T)=1.06$).
}
\end{figure}


In the left panel of Fig. \ref{fig:profile} a spatial 
configuration of the local DC velocity, 
$\omega^{i}_\mathrm{DC}=\langle\dot{\theta}_i(t)\rangle_T$, 
is shown.
Here, $\langle\dots\rangle_T$
denotes time averaging for the observation time $T$
and the resolution of DC velocity is given by $2 \pi/T$.
Note that we show the results of {\it two dimensional} systems,
which show qualitatively similar behaviors 
to three dimensional systems but show the domain structure of 
$\omega_\mathrm{DC}^i$ more clearly. 
In the right panel we show the difference in $\omega^{i}_\mathrm{DC}$,
$|\Delta \omega_\mathrm{DC}^{i,j}|
=|\omega_\mathrm{DC}^{i}-\omega_\mathrm{DC}^{j}|$,
for each bond between neighboring sites.
When $|\Delta \omega_\mathrm{DC}^{i,j}|>2\pi/T$,  phase slip occurs 
on the bond at least once during the observation time $T$.
We focus on this phase slip process
to discuss the spatial correlation of motion
instead of the direct spatial correlation of DC local velocity 
\cite{Dominguez99}.
We define such a bond as a ``disconnected'' bond.
Otherwise, if $|\Delta \omega_\mathrm{DC}|<2\pi/T$,
the bond is ``connected'' and two sites belong to the same cluster.
When both of the pair sites are pinned,
i.e., $\omega_\mathrm{DC}^i, \omega_\mathrm{DC}^j < 2\pi/T$,
$|\Delta \omega_\mathrm{DC}|$ is less than $2\pi/T$
but we regard such a bond as disconnected,
which forms nonmoving solid clusters.
We analyze a bond percolation transition by controlling driving force.
The moving solid phase is characterized by an infinitely large 
cluster, which is made from connected sites without phase slip.
The percolating phase is thus the moving solid phase.
The driving force in Fig. \ref{fig:profile} is slightly below 
the critical point and a fractal domain structure appears.

In order to perform finite size scaling,
we divide the system into subsystems,
whose linear dimension $L$ is smaller than that of a real sample 
$L_\mathrm{max}$.
Then we determine that percolation occurs
if a certain cluster reaches the two opposite sides of each subsystem.
The statistics of subsystems and samples yields percolation 
probability $P(f,L,T)$,
which monotonically grows with $f$
and becomes smaller as $T$ increases.

The reason we perform finite size scaling in terms of $L$ 
and not $L_\mathrm{max}$ as usual is the following. 
We have found that the whole system with a finite $L_\mathrm{max}$ 
eventually falls into a periodic motion above some threshold driving force. 
Therefore, for a given driving force, 
we choose to work on a system with sufficiently large $L_\mathrm{max}$ 
to eliminate this artificial periodic motion 
and analyze the connectivity of clusters 
at various length scales $L ( < L_\mathrm{max})$.

\begin{figure}
\begin{center}
\includegraphics[trim=0 260 0 -180,scale=0.23,clip]{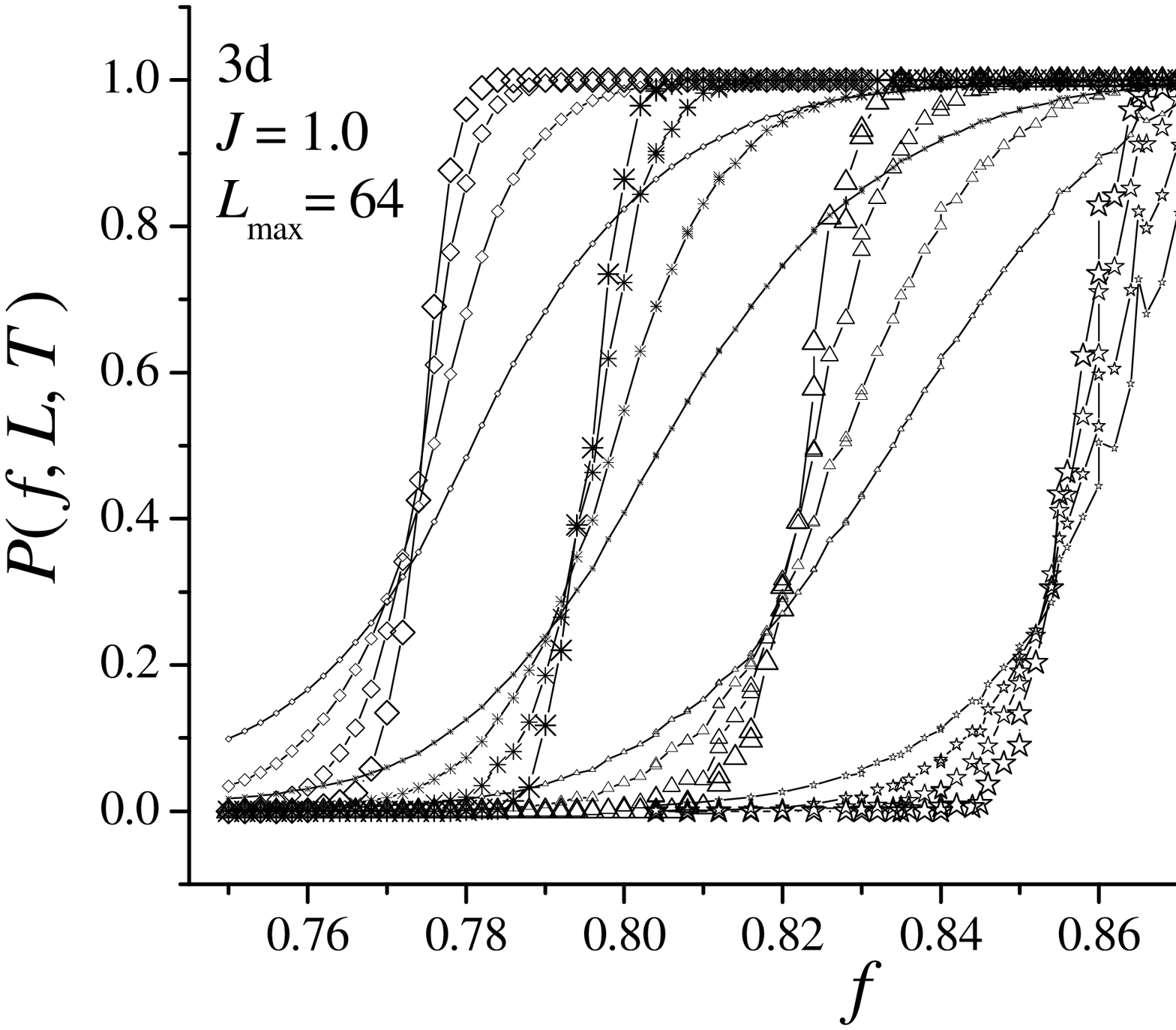}\\
\includegraphics[trim=0 250 0 -180,scale=0.23,clip]{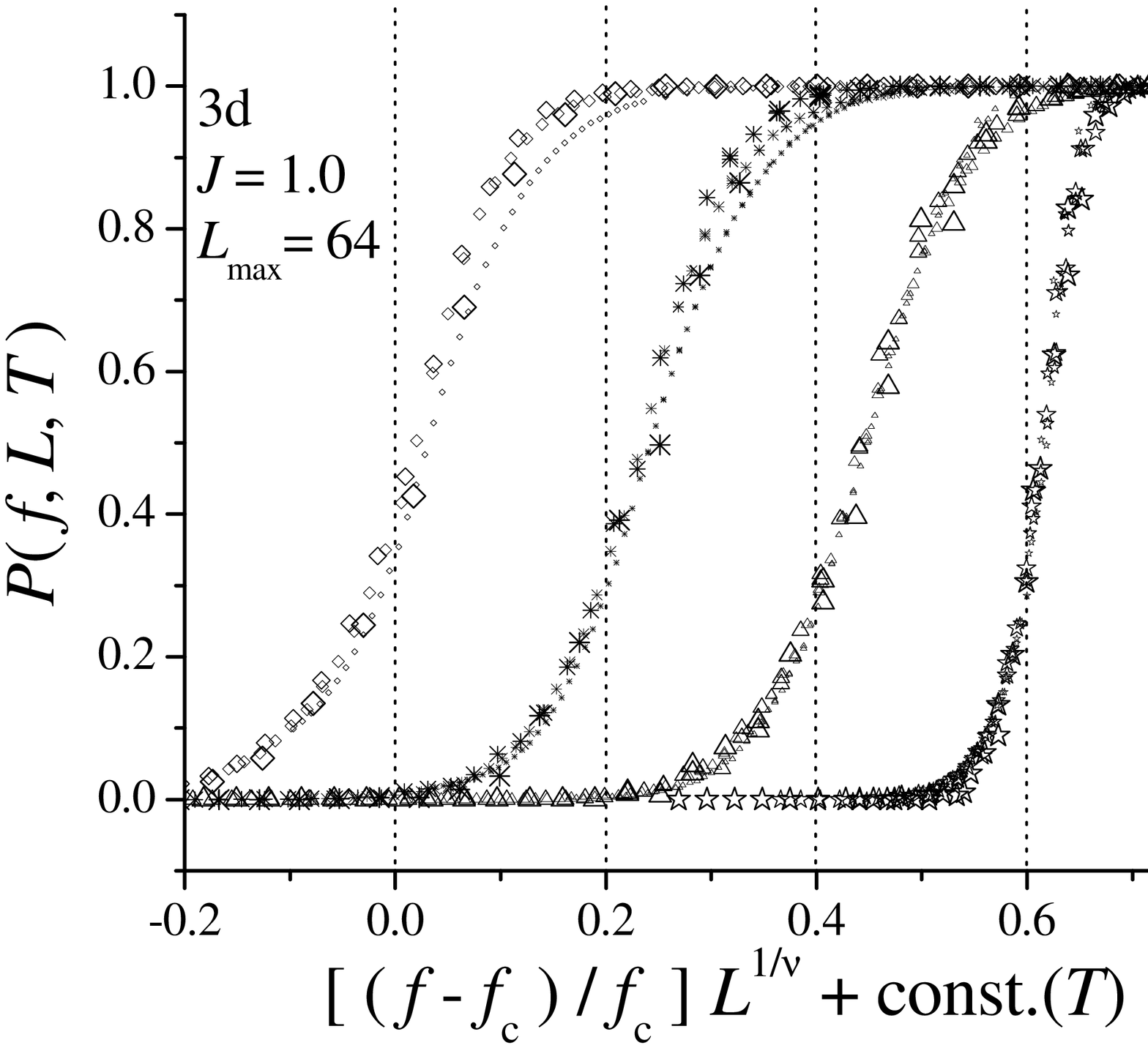}
\end{center}
\vspace{-3mm}
\caption{\label{fig:scaling}
Raw data of percolation probabilities
for various sizes and observation times (top)
and result of finite size scaling (bottom).
The x-axis of the latter is shifted by a constant which depends on $T$
and each origin is indicated by a dotted vertical line.
Result of three dimensional samples with $L_\mathrm{max}=64$
and maximum observation time $T \approx 3200$.
}
\end{figure}

In Fig. \ref{fig:scaling}, $P(f,L,T)$'s
for various $L$'s and $T$'s are plotted as functions of $f$.
We consider the large $L$ limit at fixed $T$ first
and the $T$ dependence next.
Percolation probability grows with $f$
and its shape comes closer to that of a step function of $f$
as $L$ increases.
Finite size scaling can be performed in the same way
as the stochastic percolation.
The curves for different $L$'s converge on the universal one
as the driving force is scaled as
$\tilde{f}=[ (f-f_c)/f_c ] L^{1/\nu}$
with the suitable critical force $f_c$ and the critical exponent $\nu$.
The correlation length diverges as $| (f-f_c)/f_c|^{-\nu}$.
A good conversion is obtained for each $T$ 
as shown in Fig. \ref{fig:scaling}
then $f_c(T)$'s and $\nu(T)$'s are obtained.
The fraction of connected bonds at the critical point is 0.10,
which is smaller than that for the stochastic percolation, 0.2488 
\cite{Grassberger92}, 
due to the attractive correlation of connected bonds.

\begin{figure}
\begin{center}
\includegraphics[trim=0 270 0 -190,scale=0.23,clip]{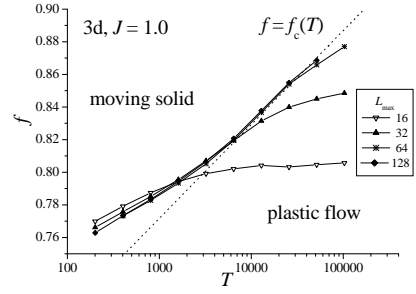}
\end{center}
\vspace{-3mm}
\caption{\label{fig:phase_diagram}
$f-T$ phase diagram.
Phase boundary between plastic flow 
and moving glass is drawn by $f=f_c(T)$.
Note that the horizontal axis is in a logarithmic scale.
The dotted line indicates $f=0.025 \ln t + 0.59$.
}
\end{figure}

The magnitude of the critical driving force for the percolation 
transition obtained from the above analysis depends on $T$.
The $T$ dependence of $f_c$ for several $L_\mathrm{max}$'s is shown
in Fig. \ref{fig:phase_diagram}.
It is expressed as 
\begin{equation}
f_c(T) = f_0 \ln( T / t_0 ),
\label{eq:fc}
\end{equation}
for a long $T$.
Here, $f_0$ and $t_0$ are constants. 
We expect that eq. (\ref{eq:fc}), 
which is consistent with the idea of shaking temperature 
\cite{Koshelev94} as mentioned later, 
holds for a sufficiently long $T$ and large $L_\mathrm{max}$ 
and the reason for the deviation between the present data and eq. (\ref{eq:fc})
for a smaller $T$ or a smaller $L_\mathrm{max}$ is considered as follows.
For $T<2000$, the $f_c$'s obtained from the simulations 
are a little larger than that expected from eq. (\ref{eq:fc}).
This discrepancy can be due to the 
existence of pinned or very low velocity sites, 
which cannot be candidates of the  percolating cluster.
They are defects of the percolation transition, 
of which effects are not taken into account in eq. (\ref{eq:fc}).
The percolation transition is not so sensitive to
the existence of such defects when they are rare and isolated.
They are not a minority, however, below $f=0.80$, 
e.g, the fractions of such sites with $\omega_\mathrm{DC}<0.050$ are
0.51, 0.090 and 0.00054 for $f$=0.75, 0.80 and 0.85, respectively. 
(Here, the local depinning threshold force \cite{Kawaguchi99,Nogawa03} 
equals 0.10.)
The critical fraction of the connected bonds
becomes larger in the lattice with such defects
and $f_c$ becomes higher.
These defects decrease rapidly with $f$
and have less effect on a higher $f_c$ for a longer $T$.

Another disagreement occurs at a longer $T$,
which is caused by the finiteness of the real sample size $L_\mathrm{max}$.
For a finite $L_\mathrm{max}$,
$f_c$ exhibits saturating behavior to a finite value above a certain $T$,
which increases with $L_\mathrm{max}$.
As mentioned before, 
this is caused by the falling of the system into limit cycle motion 
when the phase coherence length becomes comparable to $L_\mathrm{max}$.
For $T \le 51500$ ($f_c<0.870$),
the results for $L=64$ and $128$ hardly differ
and they are considered to show $f_c(T)$ for $L_\mathrm{max}=\infty$.

From these discussions eq. (\ref{eq:fc}) is expected to hold up to the 
infinite $T$.
This means that $f_c(T)$ diverges logarithmically
as $T$ approaches infinity and the moving solid {\it phase}
does not exist in the long time limit based on the present definition.
The supplemental simulation indicates that 
this behavior does not change for a system with a stronger coupling.

From another point of view, 
eq. (\ref{eq:fc}) is regarded as a type of ``phase boundary'' 
between the plastic flow and moving solid phases in the force-time plain
(See Fig. \ref{fig:phase_diagram}).
Considering the observation with fixed $f$,
the crossover time $\tau(f)$ is obtained as
\begin{equation}
\tau(f) = t_0 \exp ( f/f_0 ).
\label{eq:tau}
\end{equation}
The system behaves as if it were a moving solid
in an observation time shorter than $\tau(f)$.
Beyond this time, cracks of plastic deformation,
which are sheets of phase slip bonds,
propagate to the macroscopic scale and fluid like property is revealed.

Equation (\ref{eq:tau}) can be regarded as a thermal activation process,
$\tau \propto \exp(V/k_BT_{\mathrm {eff}})$,
if an effective temperature proportional
to $f^{-1} (\approx \omega_\mathrm{DC}^{-1})$ is supposed.
The inverse of $f_0$ is then proportional to the potential barrier $V$.
This is consistent with the idea of ``shaking temperature"
proposed by Koshelev and Vinokur
\cite{Koshelev94,Kolton99,Kolton02}.

\begin{figure}
\begin{center}
\includegraphics[trim=0 270 0 -190,scale=0.23,clip]{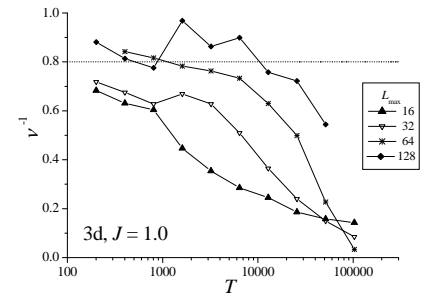}
\end{center}
\vspace{-3mm}
\caption{\label{fig:nu-t}
Observation time dependence of critical exponent.
}
\end{figure}

Next, we discuss the universality of the percolation transition
for different observation times.
In Fig.\ref{fig:nu-t}, the $T$ dependence of
the inverse of the critical exponent $\nu$ is shown.
There is a tendency for large $L_\mathrm{max}$'s, 64 and 128,
where $\nu^{-1}$ has $T$-independent value $\approx 0.8$.
The inverse of $\nu$ becomes smaller 
and approaches zero as $T$ increases.
$\nu^{-1}=0$ means that $P(f,L,T)$ does not depend on $L$,
which manifests that the system falls into the limit cycle motion, 
i.e., a finite size effect.
Such effect appears more clearly in the critical exponent than in the 
critical force, because the former is determined by the behavior in 
the whole critical regime.
So we expect the percolation transition discussed here is universal 
with respect to the observation time.

\begin{figure}
\begin{center}
\includegraphics[trim=0 270 0 -180,scale=0.23,clip]{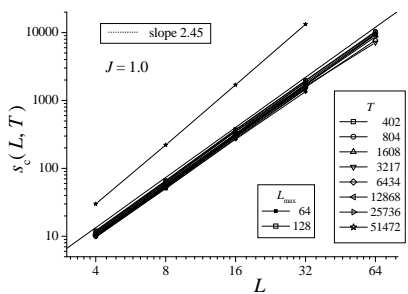}
\end{center}
\vspace{-3mm}
\caption{\label{fig:smax-l}
Relationship between maximum cluster size
and linear dimension of subsystem size 
for several observation times.
}
\end{figure}

This universality is confirmed by the property of
the percolating cluster at the critical point.
Cluster size is identified with the number of contained sites
and we define $s_c(L,T)$ as the size of the maximum cluster
in a subsystem at $f=f_c(T)$.
In Fig. \ref{fig:smax-l},
$s_c(L,T)$'s for several $T$'s
are plotted as functions of subsystem size $L$.
They are expressed as $s_c(L,T) \approx 0.35 L^{2.45}$ 
with a fractal dimension of 2.45  
and show little $T$ (or $f_c(T)$ ) dependence. 
The large deviation of the data for $L_\mathrm{max}$=64 at the longest $T$ 
is due to starting the limit cycle motion.

The universality of the transition for different $T$'s
means that the fluctuation of DC velocity has a scaleless spatial pattern, 
which depends on neither $f$ nor the phase coherence length, 
which grows with $f$, if one chooses proper time scale,
i.e., sees the DC velocity in the resolution of $2\pi/\tau(f)$.

In conclusion, we numerically investigate
the possibility of the dynamical phase transition
between plastic flow and moving solid phases, which are distinguished 
by the existence of the long range order of local DC velocity.
By analyzing the percolation of no-phase slip bonds
and its observation time dependence,
we found that the moving solid phase becomes unstable in a finite lifetime.

The condition for the connected bond,
that no phase slip occurs eternally, may seem too strict.
For example all bonds necessarily take phase slips
if thermal fluctuation exists.
It is important, however, to note that
the local symmetry 
$\lim_{T \to \infty} \Delta\omega^{i,j}_\mathrm{DC} = 0$ 
is destroyed from the beginning 
due to the random field.
The bond steadily takes a phase slip in the same direction
no matter how rarely it occurs.

The crossover time $\tau(f)$ can be defined clearly; 
the system behaves like a moving solid
in a shorter time scale than $\tau(f)$ and
macroscopic plastic deformation occurs beyond $\tau(f)$.
This crossover time increases exponentially
with the driving force.
Its characteristic scale of growth $f_0 = 0.025$ 
is very small compared with other scales such as pinning strength (=1),
therefore the crossover time increases very rapidly 
in the narrow region of $f$ 
and overcomes the macroscopic time scale.
This is a possible reason the moving solid phase is
observed in experiments.
The situation is similar to the case of 
structure glasses,
whose viscosity grows quite large and they show slow dynamics,
then it is hard to distinguish
whether an equilibrium phase transition exists.

We focused on the macroscopic plastic deformation
and distinguished between plastic flow and moving solid.
This stance is different from the conventional interest
in the liquid-crystal(Bragg glass) transition.
Although it seems natural that these transitions occurs at the same time,
the absence of the moving solid phase discussed here is not
immediately related to the absence of the long range periodic order.
For example the spatial phase order in a long span 
is possible in the plastic flow phase 
if the propagation of plastic deformation
along the domain boundary is temporally localized
and leaves no change before and after it.
On the other hand we see that
the saturation of phase correlation length to the system size
results limit cycle motion and plastic deformation is suppressed.
Then if the transition from liquid to crystal or Bragg glass,
which is not observed in the range of our simulation,
occurs at finite $f$,
the present transition would happen at the same time.

The numerical calculation was performed on the Hitachi SR8000
at the supercomputer center, ISSP, University of Tokyo
and the present study is financially supported 
by a Grant-in-Aid for Scientific Research (15540370) 
from the Japan Society for the Promotion of Science.


\end{document}